\documentstyle[12pt]{article}
\normalbaselines
\catcode `\@=11
\@addtoreset{equation}{section}

\def\section{\@startsection {section}{1}{\z@}{-3.5ex plus -1ex minus
     -.2ex}{2.3ex plus .2ex}{\normalsize\bf}}
\def\subsection{\@startsection{subsection}{2}{\z@}{-3.25ex plus -1ex minus
 -.2ex}{1.5ex plus .2ex}{\normalsize\bf}}

\def\@cite#1#2{${}^{\mbox{\scriptsize#1\if@tempswa , #2\fi}}$}
\def\thebibliography#1{\section*{References\markboth
  {REFERENCES}{REFERENCES}}\list
  {\arabic{enumi}.}{\settowidth\labelwidth{[#1]}\leftmargin\labelwidth
  \advance\leftmargin\labelsep
  \usecounter{enumi}}
  \def\newblock{\hskip .11em plus .33em minus -.07em}
  \sloppy
  \sfcode`\.=1000\relax}
 
\catcode `\@=12

\begin{document}

{\bf ON HAMILTONIAN AND QUANTUM DYNAMICS OF MASSLESS
PARTICLES.}\vspace{1.3cm}\\
\noindent
\hspace*{1in}
\begin{minipage}{13cm} Andreas Bette, \vspace{0.1cm}\\
Stockholm University, \vspace{0.1cm}\\
Department of Physics, \vspace{0.1cm}\\
Box 6730, \vspace{0.1cm}\\
S-113 85 Stockholm, Sweden.\vspace{0.1cm}\\
fax: +46-8347817 att. Andreas Bette, \vspace{0.1cm}\\
e-mail: $<$ab@vanosf.physto.se$>$

\end{minipage}

\vspace*{0.5cm}

\begin{abstract}
\noindent

A short review of special relativistic dynamics describing a particle
acted upon by an arbitrary conservative external force is presented.
If the mass of the particle is zero and the force is central then the
equations of motion turn out to be completely integrable. A well-known
result\cite{shm}.

\vskip 8pt

Hamiltonian flows on the twistor phase space \bf T \rm are constructed
which, for conservative forces and value of the helicity equal to zero,
reproduce equations of motion of the classical massless particle. For
helicities different from zero the same hamiltonian flows produce
equations of motion showing a curious "Zitterbewegung" like behaviour.

\vskip 8pt

A canonical Poincar{\'e} covariant quantization procedure on \bf T \rm
is suggested. One simple example describing a spinning and massless
3-D quantum mechanical harmonic oscillator is analysed in some detail.

\end{abstract}

\vfill
\eject

\section*{\hspace{-4mm}\hspace{2mm}INTRODUCTION AND NOTATION.}

It is possible that elementary massive particles such as electron,
proton, neutron etc.  should be regarded as bound states of a finite
number of massless and spinning interacting constituents.

\vskip 10pt

In this article a mathematical formalism with its roots in the Twistor
Theory of Penrose\cite{prmc} is investigated. The physical force in
the model is external (and conservative). Therefore its future
physical application (if any) aims only at a new phenomenological
attempt to understand where the masses of elementary particles come
from.

\vskip 10pt

The author does not claim that the model (as it now stands) describes
any known physical system.  He wishes just to show that there are some
concrete uninvestigated possibilities hidden in the mathematics of
Twistors.  The virtue of such models (when fully developed) is the
simplicity and economy of thought they provide.

\vskip 10pt

Earlier, exploring the idea of instantaneous relativistic action at a
distance\cite{wfw} in the phase space of twistors we have
shown\cite{ab1} how a free massive and spinning particle may be
thought of as a relativistic rigid rotator (endowed with intrinsic
spin) composed of two massless spinning parts.  Instantaneous refers
to the rest frame defined by the total time-like four-momentum of the
rigid rotator the two massless and spinning particles happen to
define.

\vskip 10pt

By continuing these ideas and by taking a larger number of such
massless constituents more complicated closed massive and spinnig
systems may be constructed\cite{ab2}.

\vskip 10pt

However, with the increasing number of massless parts, calculations,
although st- raightforward, become quite cumbersome.

\vskip 10pt

This fact triggered the work presented in this paper.  In order to get
an idea of how a closed system, composed of a large finite number of
massless spinning mutually interacting particles, might behave we
investigate dynamics of just one massless spinning particle acted upon
by an external conservative force.  The latter may be thought of as an
effective force coming from an inertial "source" defined by the total
freely moving composite system.

\vskip 10pt

In other words we assume that there exists a special inertial frame in
Minkowski space (the rest frame of the "infinitely" heavy "source" of
the force) to which the massless particle is bounded.

\vskip 10pt

One of the shortcomings of our approach is that there is nothing in
the formalism which tells us how to choose the external force (or
equivalently the corresponding hamiltonian) in order to describe a
physical system. Future investigations will perhaps show how the
external force should be chosen in concrete physical situations.

\vskip 10pt

The work is organized as follows:

\vskip 10pt

In the next section we review the relativistic Newton's
second law of dynamics with emphasis on the massless particle case.
It is demonstrated again that when a conservative and central force acts
on a massless particle then its relativistic equations of motion
are completely integrable\cite{shm}.

\vskip 10pt

In the third section an Hamiltonian mechanics, reproducing equations
of motion of a spinless and massless particle, is formulated on the
phase space of twistors.  Further, the approach is generalized to be
valid for non-vanishing values of the helicity.

\vskip 10pt

Finally, in the fourth section a canonical quantization is performed.
One relatively simple example, representing an analog of the 3-D
harmonic oscillator, is studied in some detail.

\vskip 10pt

Latin letters with lower case latin indices denote four-vectors and
four-tensors.  Latin letters with lower case greek indices within
brackets denote three-vectors.  In section $1$, $2$ the usual
three-vector notation (with a line over a letter) will also be used.
In section $2$ and $3$ a bar (not a line) over a letter or over an
expression denotes complex conjugation.  Lower case greek letters with
upper case latin indices (either primed or unprimed) denote spinors.
Upper case latin letters with lower case greek indices denote
twistors.  The physical units are so chosen that $c={\hbar}=1$.  The
signature of the metric $g_{ij}$ in Minkowski space is taken to be
$+---$.  The fully antisymmetric alternating four-tensor will be
denoted by $\eta_{ijkl}$.  The fully antisymmetric alternating
three-tensor will be denoted by $\epsilon_{(\alpha)(\beta)(\gamma)}$.
The usual summation convention over repeted indices will be assumed
throughout.

\section{\hspace{-4mm}\hspace{2mm}A SHORT REVIEW OF RELATIVISTIC PARTICLE
DYNAMICS.}

In this paper we intend to achieve two goals.  The first is to
describe relativistic classical dynamics of a massless spinning
particle acted upon by an external conservative force in terms of
canonical flows on the twistor phase space.  The second is to
formulate the corresponding relativistic quantum dynamics and examine
the case when the force is chosen to be of the 3-D harmonic oscillator
type.

\vskip 10pt

To define the context we first review the special relativistic version
of Newton's second law of dynamics in general and its massless limit
in particular.

\vskip 10pt

For a massive particle, Newton's second law of dynamics
may be written in the following Poincar{\'e} covariant form:

\begin{equation} {dY^{i} \over d{\tau}} = {P^{i} \over m}, \end{equation}

\begin{equation}{dP_{i} \over d{\tau}} = F_{i} \qquad \qquad where \qquad
\qquad
F^{i}P_{i}=0.\end{equation}

\vskip 10pt

$Y^{i}$ is the four-position of the particle in Minkowski space,
$P_{i}$ denotes its four- momentum, $\tau$ its proper time, $m$ its
rest mass and $F_{i}$ represents the so called four-force acting on
the particle. Suppose that the world-line of an inertial frame is
given by:

\begin{equation}X^{i} = X_{0}^{i} + (t-t_{0})t^{i},\end{equation}

\vskip 10pt

\noindent
where $X^{i}$ denotes its four-position, $t^{i}$ its constant
time-like four-velocity, $t$ its proper time and where $X_{0}^{i}$
represents a constant four-vector starting from some arbitrarily
chosen origin in Minkowski space and ending at a point (an event) on
the world-line of the inertial frame where we have put the value of
its proper time to be equal to $t_{0}$.

\vskip 15pt

Then, a space-like four-vector $r^{i}$, which represents the
particle's instantaneous \it centre of energy \rm
with respect to the inertial frame defined by (1.3), is given by:

\begin{equation}r^{i}:= R^{i} - (R^{k}t_{k})t^{i},\end{equation}

\vskip 10pt

\noindent
where

\begin{equation}R^{i} = Y^{i} - X^{i}.\end{equation}

\vskip 10pt

\noindent
Using (1.3) it yields:

\begin{equation}r^{i} = (Y^{i} - X_{0}^{i}) -
[(Y^{k}-X_{0}^{k})t_{k}]t^{i}.\end{equation}

\vskip 10pt

Let us also introduce a space-like four-vector $f_{k}$ which fulfils:

\begin{equation} f_{k}t^{k} = 0, \end{equation}

\vskip 10pt

\noindent
and represents the physical three-force, exerted on the particle under
consideration, as "measured" by an observer following the world-line of the
inertial frame, where

\begin{equation}-{f_{n}P^{n} \over P_{k}t^{k}},\end{equation}

\vskip 10pt

\noindent
represents the work done by this force. The four-force in (1.2) may
now be split into two components\cite{rw}. One along the four-velocity $t^i$
and another projected onto the space-like three-plane orthogonal to
$t^i$:

\begin{equation}F_{i}={P_{k}t^{k} \over m} f_{i} - {f_{k}P^{k} \over m}
t_{i}.\end{equation}

\vskip 15pt

The equations of motion in (1.1) and (1.2) may then be rewritten as
follows:

\begin{equation}{\dot r}^{i} = {P^{i}\over P^{k}t_{k}} - t^{i},\end{equation}

\begin{equation}{\dot P}_{i} - ({\dot P}_{k}t^{k})t_{i} = f_{i}
\qquad and \qquad {\dot P}_{i}{P}^{i}=0.\end{equation}

\vskip 10pt

The dot over a letter denotes differentiation with respect
to the proper time in the inertial frame defined by (1.3). If $m$
differs from zero this is just a reformulation of the equations given
in (1.1) and (1.2). However, in contrast to (1.1) and (1.2), the
equations in (1.10) and (1.11) are also valid for $m = 0$.
The physical three force represented by $f_{k}$ may
depend functionally on $r^{i}$ i.e.:

\begin{equation}f_{k} = f_{k}(r^{i}).\end{equation}

\vskip 10pt

The \it centre of energy \rm space-like four-vector $r^{i}$ depends on the
location of the inertial frame and also on the location of the
particle in Minkowski space.  Therefore the assumption in (1.12)
implies that the inertial frame, with the world-line given by (1.3),
is not arbitrary but constitutes a source producing the force
acting on the particle.
In the inertial frame of the source we have:

\begin{equation}t^{i}=(1, {\overline 0}),\end{equation}

\begin{equation}r^{i}=(0, {\overline r}),\end{equation}

\begin{equation}f^{i}(r^{k})=(0, {\overline f}(\overline r)),\end{equation}

\vskip 10pt

\noindent
where we have used the familiar three-vector notation.
The equations of motion in (1.10) and (1.11) now read:

\begin{equation}{\dot {\overline r}} = {{\overline p} \over E},\end{equation}

\begin{equation}{\dot {\overline p}} = {\overline f}({\overline
r}).\end{equation}

In addition we also have:

\begin{equation}m^{2}:=P^{i}P_{i} = E^{2} - {\vert {\overline p} \vert}^{2} =
E^{2} - p^{2}=constant.\end{equation}

\vskip 15pt

If the force ${\overline f}({\overline r})$ is conservative then one
has:

\begin{equation}{\overline f} = - {\nabla U}({\overline r}),\end{equation}

\vskip 10pt

\noindent
where $U({\overline r})$ represents the potential energy of the particle.
{}From (1.16)-(1.19) we obtain in the usual manner that:

\begin{equation}{\dot E} = - [{\dot {\overline r}} {\cdot } {\nabla
U({\overline r})}],\end{equation}

\vskip 10pt

\noindent
which implies the energy conservation law:

\begin{equation}H := E + U({\overline r}) = constant.\end{equation}

\vskip 10pt

The constant $H$ represents the total energy of the particle.
Differentiating (1.16), using (1.17), (1.20) and (1.21) yields:

\begin{equation}(H-U){\ddot {\overline r}} + {\nabla U({\overline r}}) - {\dot
{\overline r}}
[{\dot {\overline r}}{\cdot }{\nabla U({\overline r})}] = 0.\end{equation}

\vskip 10pt

Again,
if $m$ differs from zero and the non-relativistic condition is
fulfilled i.e.:

\begin{equation}{\vert {\dot {\overline r}} \vert} << 1,\end{equation}

\noindent
then one has that:

\begin{equation}(H-U) {\simeq } m,\end{equation}

\vskip 10pt

\noindent
which implies that the equations of
motion in (1.22) above acquire (as they of course
should) the familiar Newtonian form:

\begin{equation}m{\ddot {\overline r}} + {\nabla U({\overline r}}) {\simeq }
0.\end{equation}

\vskip 15pt

\it

Note that the relativistic non-linear equation in (1.22) is also valid
for massless particles.

\rm

\vskip 15 pt

To proceed further we assume that the force is also central i.e. that:

\begin{equation}U({\overline r}) = U({\vert {\overline r} \vert}) =
U(r).\end{equation}

\vskip 10pt

{}From (1.16), (1.17), (1.21) and (1.26) we then obtain (as in the
non-relativistic case) that the orbital angular momentum
does not change in time:

\begin{equation}{\overline L} := {\overline r} {\times } {\overline p} =
[H-U(r)] [{\overline r} {\times } {\dot {\overline r}}] =
constant,\end{equation}

\vskip 10pt

\noindent
implying plane particle motion.
${\times }$ denotes the usual
vector product in the three dimensional space (Lorentz) "perpendicular"
to the time-like direction given by $t^{i}$.

\vskip 10 pt

Choosing the space origin at the location of the
observer who follows the world-line of the inertial frame
and the $z$-axis along the three-vector ${\overline L}$ the equations
of motion, in the familiar polar coordinates on the
spatial plane perpendicular to the $z$-axis, read:

\begin{equation}{\ddot \rho} - {U'\over (H-U)}({\dot \rho}^{2} - 1)-
{L^{2} \over {{(H-U)}^{2}\rho^{3}}} = 0,\end{equation}

\begin{equation}{\dot \phi}={L \over {(H-U) \rho^{2}}} \qquad \qquad \qquad
\qquad
{\dot z} = z = 0,\end{equation}

\vskip 10pt

\noindent
where

\begin{equation}L:={{\vert {\overline L} \vert}},\end{equation}

\vskip 10pt

\noindent
and where

\begin{equation}U':={dU \over d{\rho}}.\end{equation}

\vskip 10pt

In the massless case, which is our main concern here, we have that:

\begin{equation}{\vert {\overline p} \vert}=E,\end{equation}

\vskip 10pt

\noindent
which, by the use of (1.16), implies:

\begin{equation}{{\vert {\dot {\overline r}} \vert}}=1.\end{equation}

\vskip 10pt

In the introduced polar coordinates this yields:

\begin{equation}{\dot {\rho}}^{2} + {\rho}^{2} {\dot \phi}^{2} =
1,\end{equation}

\vskip 10pt

\noindent
which inserted into the first part of (1.29) reduces the equations of
motion to a simple expression:

\begin{equation}{\dot {\rho}}^{2} = 1 - {L^{2} \over {(H-U)^{2}
{\rho}^{2}}}.\end{equation}

\vskip 10pt

The equation above may be integrated giving $t$ as a function of $\rho$:

\begin{equation}t={\pm }{{\int}
{d\rho \over {\sqrt {1 - {L^{2} \over {(H-U)^{2} \rho^{2}}}}}}},\end{equation}

\vskip 10pt

\noindent
and for the polar angle $\phi$ as a function of $\rho$ one obtains:

\begin{equation}{\phi}={\pm }L{{\int} {d\rho \over {(H-U)\rho^{2} {\sqrt {1 -
{L^{2} \over
{(H-U)^{2} \rho^{2}}}}}}}}.\end{equation}

\vskip 15pt

\it

The equations of motion, describing  massless particles acted upon by
central conservative forces, are completely integrable\cite{shm}.

\rm

\vskip 15pt

The above results may be understood in terms of symplectic (phase
space) geometry. The natural symplectic structure
on the cotangent bundle $T^*M$ (eight dimensions) of the Minkowski
space $M$ is given by:

\begin{equation} \Omega:=dP_{i}\wedge dY^{i}. \end{equation}

\vskip 10pt

The Lorentz covariant and four-translation invariant coordinates of
the momentum four-vector $P_{i}$ and the Poincar{\'e} covariant
coordinates of the position four-vector $Y^i$ regarded as functions on
$T^*M$ are canonically conjugated to each other. The ten Poincar{\'e}
covariant functions on $T^{*}M$ given by:

\begin{equation} P_{i} \qquad \qquad and \qquad \qquad M^{ij}:=2Y^{[i}P^{j]}
\end{equation}

\vskip 10pt

\noindent
define the so called momentum mapping for the action of the
Poincar{\'e} group on $T^{*}M$. In other words the algebra of their
Poisson brackets represents the commutation relations of the
Poincar{\'e } algebra.

\vskip 10pt

The particle's instantaneous position relative to the inertial source,
which is a uniformly moving point in $M$, resides on space-like planes
given by:

\begin{equation} Y^{i}t_{i}=0. \end{equation}

\vskip 10pt

If $M^{ij}$ is taken about this uniformly moving point in $M$ then the
position of the particle's \it centre of energy \rm is given by:

\begin{equation} r^{i}={M^{ij}t_{j} \over P^{m}t_{m}}. \end{equation}

\vskip 10pt

The equations of motion in (1.22) arise as a consequence of the
canonical flow in $T^{*}M$ (canonical with respect to the symplectic
structure $\Omega$ in (1.38)) generated by $H$ in (1.21) treated as a
real valued function on $T^{*}M$.

\vskip 10pt

Note that the flow generated by the mass squared function defined
in (1.18) commutes with the flow generated by the function $H$.
$m^{2}$ is therefore a constant of the physical motion.

\vskip 10pt

In the next section we are going to show that, for a massless particle,
equations of motion in (1.22) may be regarded in a completely
different way.  They will arise as a consequence of the canonical flow
generated by $H$ in (1.21) treated as a real valued function on the
twistor phase space.

\section{\hspace{-4mm}\hspace{2mm}TWISTOR PHASE SPACE FORMULATION FOR $m=0$.}

The pair $(X^{A{A^\prime}}_{0}, \ t^{A{A^\prime}})$ (here we adopt
Penrose's abstract index notation\cite{pr1}) defining the straight
world-line (of the inertial source producing the external force acting
on the massless particle) in (1.3) may be represented by two fixed
intersecting null-twistors $V^{\alpha}$, $W^{\alpha}$ in \bf T \rm (and their
twistor complex conjugates ${\bar V}_{\alpha}$, ${\bar W}_{\alpha}$)
i.e. twistors which fulfil:

\begin{equation}V^{\alpha}\bar V_{\alpha} = W^{\alpha}{\bar W}_{\alpha} =
V^{\alpha}{\bar W}_{\alpha} = 0,\end{equation}

\begin{equation}I_{\alpha \beta}V^{\alpha}W_{\beta} = {1 \over
\sqrt2},\end{equation}

\vskip 10pt

\noindent
where $I_{\alpha \beta}$ is the infinity twistor\cite{prmc,prrw}. Using the
spinor
representation this yields:

\begin{equation}V^{\alpha} = (\sigma^{A},\ \alpha_{A^\prime}),
\qquad \qquad \qquad
{\bar  V}_{\alpha} = ({\bar  \alpha_{A}},\ {\bar
\sigma}^{A^\prime}),\end{equation}

\begin{equation}W^{\alpha} = (\varsigma^{A},\ \beta_{A^\prime}),
\qquad \qquad \qquad
{\bar  W}_{\alpha} = ({\bar  \beta_{A}},\ {\bar
\varsigma}^{A^\prime}).\end{equation}

\vskip 10pt

Therefore the relation in (2.2) may be written as:

\begin{equation}I_{\alpha \beta}V^{\alpha}W^{\beta} =
\alpha^{A^\prime}\beta_{A^\prime}=
{1 \over \sqrt2},\end{equation}

\vskip 10pt

\noindent
while $X^{A{A^\prime}}_{0}$ is explicitly given by (see Ref. 2):

\begin{equation}X^{A{A^\prime}}_{0}:=i{\sqrt2}(\sigma^{A} \beta^{A^\prime}
- \varsigma^{A} \alpha^{A^\prime})=
-i{\sqrt2}({\bar  \sigma}^{A^\prime}{\bar  \beta}^{A} -
{\bar  \varsigma}^{A^\prime}{\bar  \alpha}^{A}),\end{equation}

\vskip 10pt

\noindent
and $t^{A{A^\prime}}$ by:

\begin{equation}t^{A{A^\prime}}:={\alpha^{A^\prime}{\bar  \alpha}^{A} +
\beta^{A^\prime}{\bar  \beta}^{A}}.\end{equation}

\vskip 10pt

The two fixed twistors $V$ and $W$ also define three space-like
directions rigidly attached to the inertial frame defined in (1.3):

\begin{equation}u^{a}_{(1)} \ {\equiv}\  u^{a}\ {\equiv}\ u^{A{A^\prime}}:=
\alpha^{A^\prime}{\bar  \alpha}^{A} -
\beta^{A^\prime}{\bar  \beta}^{A},\end{equation}

\begin{equation}u^{a}_{(2)} \ {\equiv}\ v^{a}\ {\equiv}\
v^{A{A^\prime}}:=\alpha^{A^\prime}{\bar  \beta}^{A} +
\beta^{A^\prime}{\bar  \alpha}^{A},\end{equation}

\begin{equation}u^{a}_{(3)}\ {\equiv}\ w^{a}\ {\equiv}\
w^{A{A^\prime}}=:i(\alpha^{A^\prime}{\bar  \beta}^{A} -
\beta^{A^\prime}{\bar  \alpha}^{A}).\end{equation}

\vskip 10pt

We may call the direction defined by $u^{a}$ the $z$-axis direction,
the direction defined by $v^{a}$ the $x$-axis direction and the
direction defined by $w^{a}$ the $y$-axis direction.

\vskip 10pt

In effect $u^{a}_{(\alpha)}$ and $t^{a}$ form a non-rotating fixed
tetrad rigidly attached to the inertial source.

\vskip 10pt

Coordinates of a variable point in \bf T \rm (i.e.  a twistor) $Z^{\alpha}$ and
coordinates of its (twistor) complex conjugate point ${\bar  Z}_{\alpha}$
will be represented by two variable Weyl spinors and their conjugates:

\begin{equation}Z^{\alpha} = (\omega^{A},\ \pi_{A^\prime}),
\qquad \qquad
{\bar  Z}_{\alpha} = ( {\bar  \pi_{A}},\ {\bar
\omega}^{A^\prime}).\end{equation}

\vskip 10pt

This spinor representation of a point in \bf T \rm is very convenient because
it shows explicitly how the Poincar{\'e} group acts on \bf T. \rm  Coordinates
of the two spinors represented by $\pi_{A^\prime}$ and $\omega^{A}$
are covariant with respect to the (identity connected part of the)
Lorentz group while four-translations $T^{a}$ act only on the
"$\omega$" spinor parts of the twistor
$Z$ and its (twistor) complex conjugate $\bar Z$
according to the
following simple rule\cite{prmc}:

\begin{equation}{\breve {\omega}}^{A}={\omega}^{A} +
iT^{A{A^\prime}}{\pi}_{A^\prime},
\qquad \qquad \qquad
{{\bar  {\breve {\omega}}}}^{A^\prime}=
{{\bar  {\omega}}}^{A^\prime} - iT^{A{A^\prime}}{\bar  \pi}_{A}.\end{equation}

\vskip 10pt

(Do not confuse $(\alpha)$ indices which label the three mutually
orthogonal physical space directions with $\alpha$ indices which label
twistors.)

\vskip 15 pt

The natural SU(2,2) invariant symplectic structure\cite{pr2,prrw,hl}  on \bf T
\rm
defines the following canonical Poisson bracket relations:

\begin{equation}\{Z^{\alpha},\ {\bar  Z}_{\beta}\} =
i{\delta}^{\alpha}_{\beta},
\qquad \qquad \qquad
\{Z^{\alpha},\ Z^{\beta}\} = \{{\bar  Z}_{\alpha},\ {\bar  Z}_{\beta}\} =
0,\end{equation}

\vskip 10pt

\noindent
which in terms of spinor variables reads:

\begin{equation}\{\omega^{A},\ {\bar  \pi_{B}}\} =
i{\delta}^{A}_{B},
\qquad \qquad \qquad
\{\pi_{B^\prime},\ {\bar  \omega}^{A^\prime}\} =
i{\delta}^{A^\prime}_{B^\prime},\end{equation}

\begin{equation}\{\omega^{A},\ \omega^{B}\} = \{\omega^{A},\ \pi_{A^\prime}\}=
\{\pi_{A^\prime},\ \pi_{B^\prime}\} =
\{\pi_{A^\prime},\ {\bar  \pi}_{B}\} = 0,\end{equation}

\begin{equation}\{{\bar  \omega}^{A^\prime},\ {\bar  \omega}^{B^\prime}\}=
\{{\bar  \omega}^{A^\prime},\ {\bar  \pi_{A}}\} =
\{{\bar  \pi_{A}},\ {\bar  \pi_{B}}\} =
\{{\omega}^{A},\ {\bar  \omega}^{B^\prime}\} = 0.\end{equation}

\vskip 15pt

The linear
four-momentum $P_{a}$ and the angular four-momentum
${  M}_{ab} = - {  M}_{ba}$
of a massless particle may be regarded as given\cite{prmc}
by the following set of Poincar{\'e} covariant functions on \bf T: \rm

\begin{equation}P_{a}:= \pi_{A^\prime}{\bar  \pi}_A,\end{equation}

\begin{equation}{  M}_{ab}:= i{\bar
\omega}_{(A^\prime}{\pi}_{B^{\prime})}{\epsilon}_{AB}-i{\omega}_{(A}{\bar
\pi}_{B)} {\epsilon}_{A^{\prime}B^{\prime}}.\end{equation}

\vskip 10pt

The canonical Poincar{\'e} covariant
Poisson brackets in (2.13)-(2.16) imply that $P_{a}$ and
${  M}_{ab}$ in (2.17)-(2.18) fulfil the Poisson bracket
relations of the Poincar{\'e} algebra\cite{hl}:

\begin{equation}\{P_{a},\ P_{b}\}=0,\end{equation}

\begin{equation}\{{  M}_{ab},\ P_{c}\}=2g_{c[a}P_{b]},\end{equation}

\begin{equation}\{{  M}_{ab},\ {  M}_{cd}\}=
2(g_{c[a}{  M}_{b]d}) + g_{d[b}{  M}_{a]c}).\end{equation}

\vskip 10pt

The Poisson bracket relations in (2.19) - (2.21) define the momentum
mapping for the action of the Poincar{\'e} group on \bf T. \rm

\vskip 10pt

The Pauli-Luba{\'n}ski spin
four-vector of a massless particle:

\begin{equation}S^{a}:={1 \over 2}{\eta}^{abcd}P_{b}{  M}_{cd}\end{equation}

\vskip 10pt

\noindent
reduces itself to a simple expression\cite{prmc}
(use spinor representation\cite{prrw}
of ${\eta}^{abcd}$ to prove it):

\begin{equation}S^{a}=sP^{a},\end{equation}

\vskip 10pt

\noindent
where

\begin{equation}s = {1 \over2}
(Z^{\alpha}\bar  Z_{\alpha}) =
{1 \over2}({\omega^{A}}{\bar  \pi_{A}} +
{\pi_{A^\prime}}{\bar  \omega^{A^\prime}}).\end{equation}

\vskip 10pt

The kinetic energy of the massless particle, in the special inertial
frame defined by the source, will be defined by the following
function on \bf T. \rm

\begin{equation}E := \pi_{C^{\prime}}{\bar
\pi}_{C}t^{C{C^\prime}},\end{equation}

\vskip 10pt

\noindent
while the linear three-momentum will be given by:

\begin{equation}p_{(\alpha)} := -\pi_{C^{\prime}}{\bar
\pi}_{C}u^{C{C^\prime}}_{(\alpha)}.\end{equation}

\vskip 10pt

If ${M}_{ab}$ is taken about the inertial source producing the force
(i.e.  about a uniformly moving point in Minkowski space $M$) then the
functions representing angular momentum of the massless particle about
this source, are given by:

\begin{equation}J_{(\alpha)} := {1 \over 2}\epsilon_{(\alpha)(\gamma)(\delta)}
{  M}_{ab}u_{(\gamma)}^{a}u_{(\delta)}^{b},\end{equation}

\vskip 10pt

\noindent
where

\begin{equation}\epsilon_{(\alpha)(\beta)(\gamma)}:=\eta_{abcd}t^{a}u^{b}_{(\alpha)}
u^{c}_{(\beta)}u^{d}_{(\gamma)}.\end{equation}

\vskip 10pt

\noindent
while the three functions representing the position of the \it centre
of energy \rm relative to the inertial source are given by:

\begin{equation}{  y}_{(\alpha)}:=-{{  M}_{ab}t^{b}u^{a}_{(\alpha)}
\over P_{c}t^{c}}.\end{equation}

\vskip 10pt

Explicitly for the three components of the total angular momentum
it yields:

\begin{equation}J_{z}=J_{(1)}=-[{1 \over \sqrt{2}}
({\bar \alpha}_{A}{\bar \beta}^{B} +
{\bar \beta}_{A}{\bar \alpha}^{B}){\bar \pi}_{B}\omega^{A}
+ {1 \over \sqrt{2}}
(\alpha_{A^\prime}\beta^{B^\prime} +
\beta_{A^\prime}\alpha^{B^\prime})\pi_{B^\prime}{\bar \omega}^{A^\prime}],
\end{equation}

\vskip 10pt

\begin{equation}J_{x}=J_{(2)}={1 \over \sqrt{2}}
({\bar \alpha}_{A}{\bar \alpha}^{B}-{\bar \beta}_{A}{\bar \beta}^{B})
{\bar \pi}_{B}\omega^{A}
+{1 \over \sqrt{2}}
(\alpha_{A^\prime}\alpha^{B^\prime}-\beta_{A^\prime}\beta^{B^\prime})
\pi_{B^\prime}{\bar \omega}^{A^\prime},\end{equation}

\vskip 10pt

\begin{equation}J_{y}=J_{(3)}={-i \over \sqrt{2}}
({\bar \alpha}_{A}{\bar \alpha}^{B} + {\bar \beta}_{A}{\bar \beta}^{B})
{\bar \pi}_{B}\omega^{A}
+{i \over \sqrt{2}}
(\alpha_{A^\prime}\alpha^{B^\prime} + \beta_{A^\prime}\beta^{B^\prime})
\pi_{B^\prime}{\bar \omega}^{A^\prime},\end{equation}

\vskip 10pt

\noindent
and for the three components of the position vector of the  \it centre
of energy \rm one obtains:

\begin{equation}{z}=y_{(1)}=
{-i \over \pi_{C^{\prime}}{\bar  \pi}_{C}t^{C{C^\prime}}}
[{1 \over \sqrt{2}}
({\bar \alpha}_{A}{\bar \beta}^{B} +
{\bar \beta}_{A}{\bar \alpha}^{B}){\bar \pi}_{B}\omega^{A}
 -{1 \over \sqrt{2}}
(\alpha_{A^\prime}\beta^{B^\prime} +
\beta_{A^\prime}\alpha^{B^\prime})\pi_{B^\prime}{\bar \omega}^{A^\prime}],
\end{equation}

\vskip 10pt

\begin{equation}{x}=y_{(2)}=
{-i \over \pi_{C^{\prime}}{\bar  \pi}_{C}t^{C{C^\prime}}}
[{1 \over \sqrt{2}}
({\bar \beta}_{A}{\bar \beta}^{B}-{\bar \alpha}_{A}{\bar \alpha}^{B})
{\bar \pi}_{B}\omega^{A}
-{1 \over \sqrt{2}}
(\beta_{A^\prime}\beta^{B^\prime}-\alpha_{A^\prime}\alpha^{B^\prime})
\pi_{B^\prime}{\bar \omega}^{A^\prime}],\end{equation}

\vskip 10pt

\begin{equation}{y}=y_{(3)}=
{1 \over \pi_{C^{\prime}}{\bar  \pi}_{C}t^{C{C^\prime}}}
[{1 \over \sqrt{2}}
({\bar \alpha}_{A}{\bar \alpha}^{B} + {\bar \beta}_{A}{\bar \beta}^{B})
{\bar \pi}_{B}\omega^{A}
+{1 \over \sqrt{2}}
(\alpha_{A^\prime}\alpha^{B^\prime} + \beta_{A^\prime}\beta^{B^\prime})
\pi_{B^\prime}{\bar \omega}^{A^\prime}].\end{equation}

\vskip 10pt

The helicity $s$ in (2.24) is a conformal scalar and thereby also a
Poincar{\'e} scalar function on \bf T. \rm  Therefore the function $s$ Poisson
commutes with all the functions introduced in (2.30)-(2.35).

\vskip 10pt

Using (2.19)-(2.21) one obtains following Poisson bracket relations:

\begin{equation}\{{  y}_{(\alpha)},\ {  y}_{(\beta)}\} =
{{s \epsilon_{(\alpha)(\beta)(\gamma)}p_{(\gamma)}} \over E^3},\end{equation}

\begin{equation}\{p_{(\beta)},\ {  y}_{(\alpha)}\} =
\delta_{(\alpha)(\beta)},\end{equation}

\begin{equation}\{E,\ {  y}_{(\alpha)}\} = {p_{(\alpha)} \over
E},\end{equation}

\begin{equation}\{E,\ p_{(\beta)}\} = \{E,\ J_{(\alpha)}\} = 0,\end{equation}

\begin{equation}\{J_{(\alpha)}, \ J_{(\beta)}\}=
\epsilon_{(\alpha)(\beta)(\gamma)}J_{(\gamma)},\end{equation}

\begin{equation}\{J_{(\alpha)}, \ {  y}_{(\beta)}\} =
\epsilon_{(\alpha)(\beta)(\gamma)}
{  y}_{(\gamma)},\end{equation}

\begin{equation}\{J_{(\alpha)}, \ p_{(\beta)}\} =
\epsilon_{(\alpha)(\beta)(\gamma)}
p_{(\gamma)}.\end{equation}

\vskip 10pt

The above commutation relations are quite reasonable from the physical
point of view. Apart from (2.36) they are what one should expect.

\vskip 10pt

The energy $H$ in (1.21) (for $m=0$) may now be treated as a function
on \bf T. \rm The canonical flow which $H$ generates on \bf T \rm is then
explicitly
given by:

\begin{equation}\dot \omega^{A} = \{H, \ \omega^{A}\} = -i{{\partial H} \over
{\partial
{{\bar  \pi}_{A}}}},\end{equation}

\begin{equation}{\dot \pi}_{B^\prime}= \{H, \ \pi_{B^\prime}\} =
- i {{\partial H} \over {\partial {\bar  \omega}^{B^\prime}}}.\end{equation}

\vskip 10pt

For functions describing physical variables in (2.27)-(2.35)
it yields:

\begin{equation}{\dot {  y}}_{(\alpha)} =\{H,\ {  y}_{(\alpha)}\} =
\{E+U,\ {  y}_{(\alpha)}\}= {p_{(\alpha)} \over E}-
{{s \epsilon_{(\alpha)(\beta)(\gamma)}p_{(\gamma)}} \over E^3}
{{\partial U} \over {\partial {  y}_{(\beta)}}} ,\end{equation}

\begin{equation}{\dot p}_{(\alpha)} = \{E+U,\ p_{(\alpha)}\} = \{U,\
p_{(\alpha)}\}
= {{\partial U} \over {\partial {  y}_{(\beta)}}}
\{{  y}_{(\beta)},\ p_{(\alpha)}\}
= - {{\partial U} \over {\partial {y}_{(\alpha)}}},\end{equation}

\vskip 10pt

\noindent
which for $s=0$ implies the equations of motion in (1.22). For $s$ not
being equal to zero the above description is a generalization of the
massless phase space dynamics. Due to (2.36) the velocity of the \it
centre of energy \rm and the velocity of the massless spinning
particle cease to define the same physical quantity. This indicates
the extended nature of the massless spinning particle. One possible
interpretation of this Zitterbewegung-like behaviour is presented in
Ref. 11. Another is that, when external forces are acting on a spinning
massless particle, the direction of motion of the centre of its \it
entire \rm kinetic energy (including energy generated by the helicity)
simply ceases to be parallell with the (null-) direction of motion of
the centre of its \it translational \rm kinetic energy.
Relative to the rest frame of the source the velocity of the centre of
\it entire \rm kinetic energy of the interacting massless spinning
particle exceeds the velocity of light.  The velocity of the centre
of its translational kinetic energy remains on the other hand always
equal to the velocity of light.  The two velocities coincide when the
massless spinning particle is moving freely.

\vskip 10pt

Using arguments having their origins in symplectic geometry we will
now demonstrate that for central forces, irrespective whether the
helicity is equal to zero or not, equations in (2.45)-(2.46) are
completely integrable.

\vskip 10pt

First we note that on the 8-D twistor phase space \bf T \rm the energy
function $H$ Poisson commutes with the helicity function $s$. For a
fixed value of $s$ the energy function $H$ may therefore be regarded
as a function on the 6-D phase space obtained as a reduction of \bf T \rm
by the function $s$. In other words, points in the reduced 6-D phase
space are represented by these curves on \bf T \rm of the hamiltonian
flow generated by $s$ which lie on the shell-surface given by a fixed
value of the function $s$.

\vskip 10pt

In the same way components of the total angular momentum $J_{(\alpha)}$
which Poisson commute with the helicity function $s$ on \bf T \rm
may be regarded as functions on the reduced 6-D phase space.
For central forces they also commute with the energy function $H$.
So they also commute with $H$ on the reduced 6-D phase space.

\vskip 10pt

Any of the components of the total angular momentum $J_{(\alpha)}$ say
$J_{(1)}$, the square of the total angular momentum
$J^{2}:=J_{(\alpha)}J_{(\alpha)}$ and $H$ are mutually Poisson
commuting functions on the 6-D reduced phase space

\vskip 10pt

Reduction of the 6-D phase space by the two mutually Poisson commuting
functions $J_{(1)}$ and $J^{2}$ produces, for each fixed value of these
functions, a 2-D phase space. The energy function $H$ may now be regarded
as a function on this 2-D phase space. But
equations of motion on a 2-D phase space
are always completely integrable.

\vskip 10pt

The somewhat unexpected result\cite{shm} reproduced in the
introduction, proving the complete integrability of the equations of
motion of a massless and spinless particle moving under the action
of a conservative and central force, thus turns out to be a special
case of the general feature as described above.

\vskip 10pt

Consequently, for external central forces, there exists a general solution of
the equations of motion in (2.45)-(2.46). For $s=0$ this solution
reduces itself to that presented in (1.36)-(1.37). To find the general
solution by means of quadratures we proceed as follows. In the rest
frame of the source we reintroduce the standard vector notation.

\vskip 10pt

We recall that the translational kinetic energy of the massless
spinning particle is denoted by:

\begin{equation}
E=\vert {\overline p} \vert,
\end{equation}

\vskip 10pt

\noindent
the distance from the source to the centre of the
entire kinetic energy of the particle by:

\begin{equation}
r=\vert {\overline r} \vert=\sqrt{{\rho}^{2}+z^{2}}
\end{equation}

\vskip 10pt

\noindent
($\rho$ and $z$ are plane polar coordinates of the corresponding
position vector) and the total energy of the massless and spinning
particle by:

\begin{equation}
H=E+U(r).
\end{equation}

\vskip 10pt

For later convenience we introduce a function defined by:

\begin{equation}
f(r):=-r{dU \over dr},
\end{equation}

\vskip 10pt

\noindent
while we also note that in standard vector notation one has:

\begin{equation}
{\overline J}=
{\overline r} \times {\overline p} + {s{\overline p} \over E}
\end{equation}

\vskip 10pt

\noindent
where ${\overline J}$ is the total angular momentum vector of the
massless spinning particle and where ${\overline p}$ is its linear momentum
vector.

\vskip 10pt

In three vector notation the equations of motion in (2.45)-(2.46)
now read:

\begin{equation}
{\dot {\overline p}}={f(r) \over r^{2}}{\overline r}
\end{equation}

\begin{equation}
{\dot {\overline r}}={{\overline p} \over E}
- {s \over E^{3}}({\overline r} \times {\overline p})
{f(r) \over r^{2}}
\end{equation}

\vskip 10pt

Standard calculations also give:

\begin{equation}
({\overline r} \cdot {\overline p})^{2}=E^{2}r^{2}+s^{2}-J^{2}.
\end{equation}

\vskip 10pt

Choosing the direction of the constant total angular momentum
along the positive direction of the $z$ axis implies:

\begin{equation}
{\overline J}=
J{\overline e}_{z}.
\end{equation}

\vskip 10pt

\noindent
(2.55), (2.51) and (2.54) then yield:

\begin{equation}
z=\pm {rs \over J} \sqrt{1+{s^{2}-J^{2} \over E^{2}r^{2}}}.
\end{equation}

\vskip 10pt

{}From (2.52) follows that:

\begin{equation}
{\dot E}=
{f(r)({{\overline r} \cdot {\overline p}}) \over r^{2}E},
\end{equation}

\vskip 10pt

\noindent
while from (2.49) follows that:

\begin{equation}
{\dot E}=
-{dU \over dr}{\dot r}.
\end{equation}

\vskip 10pt

Using (2.50), (2.54), (2.57), (2.58) we finally obtain:

\begin{equation}
{\dot r}=\sqrt{1+{s^{2}-J^{2} \over E^{2}r^{2}}}.
\end{equation}

\vskip 10pt

For $s=0$ the above results imply that $z=0$ and that (2.59) is
equivalent to (1.36).

\vskip 10pt

Using (2.51), (2.55) one obtains:

\begin{equation}
p_{\varphi}={{E^{2} \rho J} \over {s^{2} + r^{2} E^{2}}},
\end{equation}

\begin{equation}
p_{z}={{EJ(s^{2} +z^{2}E^{2})} \over {s(s^{2} + r^{2} E^{2})}},
\end{equation}

\begin{equation}
p_{\rho}={{E^{3} J z \rho} \over {s(s^{2} + r^{2} E^{2})}}.
\end{equation}

\vskip 10pt

The above results and the equation of motion in (2.53) yield:

\begin{equation}
{\dot \varphi}={EJ \over {s^{2}+r^{2}E^{2}}}(1+{s^{2}f(r) \over r^{2}E^{2}})
\end{equation}

\vskip 10pt
\noindent
which for $s=0$ imply (1.37).

\vskip 10pt

For specific choices of the potential energy in (2.49) the above
general equations may easily be integrated and plotted by the use of
modern computer programs such as e.g.  Maple V, Release 3.  The
concrete results of such calculations we hope to be able to present in
a future publication.


\section{\hspace{-4mm}\hspace{2mm}MASSLESS RELATIVISTIC QUANTUM DYNAMICS.}

In this section a quantization of the twistor phase space dynamics is
suggested.  The energy eigenvalue equation corresponding to the
potential of the 3-D harmonic oscillator is studied in some detail.

\vskip 10 pt

A non-standard, as opposed to the standard procedure introduced by
Penrose\cite{prmc}, canonical twistor quantization is obtained by
means of a natural prescription {\'a} la Dirac\cite{dpam1,dpam2}
given by:

\begin{equation}{\hat \omega}^{A} :=
- {{\partial} \over {\partial {\bar \pi}_{A}}}, \qquad \qquad
{\hat {\bar \omega}}^{A^\prime} := {{\partial} \over {\partial
{\pi}_{A^\prime}}},\end{equation}

\begin{equation}{\hat {\bar \pi}}_{A} := {\bar \pi}_{A}, \qquad \qquad
{\hat {\pi}}_{A^\prime} := {\pi}_{A^\prime}.\end{equation}

\vskip 10pt

The Poisson brackets relations in (2.14) - (2.16) will hereby be
replaced by the corresponding commutators turning the classical
twistor phase space dynamics of a massless particle into its quantum
mechanical analog.

\vskip 10pt

So by the use of (3.1)-(3.2) the linear four-momentum functions in
(2.17), the angular four-momentum functions in (2.18), the helicity
function in (2.24) turn into the corresponding operators:

\begin{equation}{\hat P}_{a}:={\bar \pi}_{A}{\pi}_{A^\prime},\end{equation}

\begin{equation} {\hat M}^{ab}:=
i{\pi}^{({A^\prime}}
{{\partial} \over {\partial
{\pi}_{{B^\prime})}}}\epsilon^{AB}
+
i{\bar \pi}^{({A}}{{\partial} \over {\partial {\bar \pi}_{{B)}}}}
\epsilon^{{A^\prime}{B^\prime}},
\end{equation}

\begin{equation}{\hat s}: =
-{1 \over 2}
({\bar \pi}_{A}{\partial \over \partial {\bar \pi}_{A}}-
\pi_{A^\prime}{\partial \over \partial \pi_{A^\prime}}).\end{equation}

\vskip 10pt

The Poisson bracket relations in (2.19)-(2.21) ensure that operators
in (3.3) and (3.4) obey commutation relations of the Poincar{\'e}
algebra.
In this sense the quantization suggested in (3.1)-(3.2) is
Poincar{\'e} covariant.  It is not conformally covariant, however.
Besides,
the helicity operator in (3.5)
commutes with all the operators
in (3.3) and (3.4).

\vskip 10pt

Reassuming, the helicity function $s$ in (2.24), the three
functions representing components of the angular momentum in
(2.30)-(2.32), the three functions representing components of the
position vector of the \it centre of entire energy \rm in (2.33)-(2.34) turn
into linear differential operators acting on the infinitely
dimensional space $\Gamma$ of complex valued functions defined on the
(two complex dimensional) configuration space $\Pi$ spanned by Weyl
spinors (multiplicative operators in (3.2) "acting" on $\Gamma$
itself). It is obvious from (3.3) that the functions in (2.25) -
(2.26) representing the energy and the linear three-momentum of the
massless particle turn into multiplicative operators "acting" on
$\Gamma$. The Hamiltonian function $H$ in (1.21) becomes a (possibly
non-linear and non-local) differential operator acting on $\Gamma$.
(For free particles $H$ is simply the kinetic energy in (2.25) i.e. a
multiplicative operator).

\vskip 10pt

A Poincar{\'e} invariant scalar product on the space of complex valued
functions on $\Pi$ will, tentatively, be defined as:

\begin{equation}<g_{1}\vert g_{2}>:=
\int
[{\bar g}_{1}({\bar \pi}_{B}, \ \pi_{B^\prime})
g_{2}(\pi_{B^\prime}, \ {\bar \pi}_{B})]
d\pi^{A^\prime}\wedge
d\pi_{A^\prime} \wedge d{\bar \pi}^{A} \wedge d{\bar \pi}_{A}.\end{equation}

\vskip 10pt

The subspace ${\aleph}$ of $\Gamma$ consisting of functions having
finite norm (with respect to the scalar product in (3.6)) defines a
Hilbert space of quantum states of a massless particle.

\vskip 10pt

Note that, provided state functions vanish at infinity i.e. vanish for
infinite values of the kinetic energy (see 3.9), the operators in
(3.3) and (3.4) are hermitian with respect to the scalar product in
(3.6). Our tentative choice of the scalar product is based on this
fact.

\vskip 10pt

The helicity operator in (3.5) may be regarded as "hermitian" with
respect to the scalar product in (3.6) if the two functions $g_{1}$
and $g_{2}$ in (3.6) represent eigenstates corresponding to the same
eigenvalue of the helicity operator in (3.5).

\vskip 10pt

A point in the four-dimensional configuration space $\Pi$ has a
relatively clear physical interpretation. Three of its coordinates
combine to give components of the linear null four-momentum (the pole)
while the fourth coordinate represents the phase (the
flag) of the Weyl spinor.

\vskip 10pt

This implies that instead of $\pi_{A^\prime}$ and ${\bar \pi}_{A}$ we
may choose more physical coordinates on ${\Pi}$ to label its points.
These physical coordinates are $E=p$ - the kinetic energy of the massless
quantum particle as "observed" in the inertial frame of the source
moving along the world-line in (1.3), $\varphi$ -, $\theta$ - the
angles denoting the direction of motion of the particle in this frame
and $\psi$ - the angle representing the phase variable in this frame.
The correspondence between the two types of coordinates is given by:

\begin{equation}
{\bar \alpha}^{A}{\bar \pi}_{A} =
\pm {\sqrt E} e^{{i\varphi \over 2}}
e^{-i\psi}\sin{\theta \over 2},\qquad  \qquad
\alpha^{B^\prime}\pi_{B^\prime}=
\pm {\sqrt E} e^{-{i\varphi \over 2}}
e^{i\psi}\sin{\theta \over 2},\end{equation}

\begin{equation}{\bar \beta}^{B}{\bar \pi}_{B}=
\pm {\sqrt E} e^{-{i\varphi \over 2}}
e^{-i\psi}\cos{\theta \over 2}, \qquad \qquad
\beta^{A^\prime}\pi_{A^\prime}=
\pm {\sqrt E} e^{{i\varphi \over 2}}
e^{i\psi}\cos{\theta \over 2},\end{equation}

\vskip 10pt

\noindent
and inversely by:

\begin{equation}
E=({\alpha}^{B^\prime}{\bar \alpha}^{B} +
{\beta}^{B^\prime}{\bar \beta}^{B}){\pi}_{B^\prime} {\bar \pi}_{B},
\end{equation}

\begin{equation}e^{4i\psi}=
{({\alpha}^{A^\prime}{\pi}_{A^\prime})
({\beta}^{B^\prime}\pi_{B^\prime})
\over
({\bar \beta}^{A}{\bar \pi}_{A})({\bar \alpha}^{B}{\bar \pi}_{B})},
\end{equation}

\begin{equation}
e^{2i\varphi}=
{{p_{(2)} + ip_{(3)}} \over {p_{(2)} - ip_{(3)}}}=
{({\bar \alpha}^{A}{\bar \pi}_{A})
(\beta^{A^\prime}\pi_{A^\prime})
\over
({\bar \beta}^{B}{\bar \pi}_{B})
(\alpha^{B^\prime}\pi_{B^\prime})},\end{equation}

\begin{equation}\cos\theta= {p_{(1)} \over E}=
{(\beta^{A^\prime}{\bar \beta}^{A} -
\alpha^{A^\prime}{\bar \alpha}^{A})
\pi_{A^\prime} {\bar \pi}_{A}
\over
(
\alpha^{B^\prime}{\bar \alpha}^{B} +
\beta^{B^\prime}{\bar \beta}^{B}
)
\pi_{B^\prime} {\bar \pi}_{B}}.\end{equation}

\vskip 10pt

{}From (3.4) and (2.27), (2.30)-(2.32) it follows that the hermitian
differential operators corresponding to the three components of the
the total angular momentum $J_{(1)}$, $J_{(2)}$, $J_{(3)}$ are given
by:

\begin{equation}{\hat J}_{(1)}=
{1 \over \sqrt{2}}
({\bar \alpha}_{A}{\bar \beta}^{B}
+
{\bar \alpha}^{B}{\bar \beta}_{A}){\bar \pi}_{B}
{\partial \over \partial {\bar \pi}_{A}}
- {1 \over \sqrt{2}}
(\alpha_{A^\prime}\beta^{B^\prime}
+
\beta_{A^\prime}\alpha^{B^\prime})\pi_{B^\prime}
{\partial \over \partial \pi_{A^\prime}},\end{equation}

\begin{equation}{\hat J}_{(2)}=
{1 \over \sqrt{2}}
({\bar \beta}_{A}{\bar \beta}^{B}-{\bar \alpha}_{A}{\bar \alpha}^{B})
{\bar \pi}_{B} {\partial \over \partial {\bar \pi}_{A}}
- {1 \over \sqrt{2}}
(\beta_{A^\prime}\beta^{B^\prime}-\alpha_{A^\prime}\alpha^{B^\prime})
\pi_{B^\prime}{\partial \over \partial \pi_{A^\prime}},\end{equation}

\begin{equation}
{\hat J}_{(3)}=
{i \over \sqrt{2}}
({\bar \alpha}_{A}{\bar \alpha}^{B} + {\bar \beta}_{A}{\bar \beta}^{B})
{\bar \pi}_{B} {\partial \over \partial {\bar \pi}_{A}}
+{i \over \sqrt{2}}
(\alpha_{A^\prime}\alpha^{B^\prime} + \beta_{A^\prime}\beta^{B^\prime})
\pi_{B^\prime}{\partial \over \partial \pi_{A^\prime}}.
\end{equation}

\vskip 10pt

The hermitian differential operator representing the square of the
absolute value of the total angular momentum:

\begin{equation}
{\hat J}^{2}:=
{\hat J}_{(1)}{\hat J}_{(1)}+
{\hat J}_{(2)}{\hat J}_{(2)}+
{\hat J}_{(3)}{\hat J}_{(3)}
\end{equation}

\vskip 10pt

\noindent
may, as well-known, be written as:

\begin{equation}
{\hat J}^{2}=({\hat J}_{(2)} + i{\hat J}_{(3)})
({\hat J}_{(2)} - i {\hat J}_{(3)}) +{\hat J}_{(1)}{\hat J}_{(1)}
-{\hat J}_{(1)}.
\end{equation}

\vskip 10pt

For the massless harmonic oscillator the classical (total energy)
Hamilton function on \bf T \rm is given by:

\begin{equation}H=E+{k \over 2}r^{2}
\qquad \qquad
r^{2}:=
y_{(\alpha)}y_{(\alpha)}
\end{equation}

\vskip 10pt

\noindent
where $y_{(1)}$, $y_{(2)}$, $y_{(3)}$ are functions
on \bf T \rm  in accordance with
(2.33)-(2.35).  Using the quantization prescription in (3.1) and (3.2)
(and the principle of normal ordering) these functions may easily be turned
into differential operators acting on $\Gamma$.  However,
from (2.54) we already know
that on \bf T \rm one has:

\begin{equation}
r^{2}=y_{(\alpha)}y_{(\alpha)}=
{{J^{2}-s^{2} + {({  y}_{(\alpha)}p_{(\alpha)})}^{2}}
\over E^{2}}.\end{equation}

\vskip 10pt

Direct calculations also show that in terms of twistor variables we have:

\begin{equation}
-i{\kappa}:=y_{(\alpha)}p_{(\alpha)}=
{i \over 2} {({\bar \pi}_{A}
{\omega}^{A} - \pi_{A^\prime} {{\bar \omega}}^{A^\prime})}.
\end{equation}

\vskip 10pt

Therefore normal ordering of terms yields:

\begin{equation}{\hat \kappa}:=
{1 \over 2}
{({\bar \pi}_{A}
{\partial \over \partial {\bar \pi}_{A}}
+
\pi_{A^\prime}
{\partial \over \partial \pi_{A^\prime}}
+4)}\end{equation}

\vskip 10pt

\noindent
which implies that:

\begin{equation}[{\hat \kappa}, \ \pi_{A^\prime}{\bar \pi}_{A}]=
\pi_{A^\prime}{\bar \pi}_{A}.\end{equation}

\begin{equation}[{\hat \kappa}, \ t^{AA^\prime}\pi_{A^\prime}{\bar \pi}_{A}]=
[{\hat \kappa}, \ E]=t^{AA^\prime}
\pi_{A^\prime}{\bar \pi}_{A}=E.\end{equation}

\vskip 10pt

\noindent
where by square brackets we denote the commutator.

\vskip 10pt

(3.20)-(3.21),
normal ordering of terms and quantization prescription in
(3.1)-(3.2) imply  that the hermitian differential
operator corresponding to the function $r^{2}$ in (3.18)-(3.19) is
given by:

\begin{equation}{\hat r}^{2}
:=
{1 \over E^{2}}
({\hat J}^{2}-{\hat s}^{2} -
{\hat \kappa}^{2}).\end{equation}

\vskip 10pt

For the quantum mechanical harmonic oscillator the Hamilton
(total energy) hermitian differential operator acting on $\Gamma$ is
represented by:

\begin{equation}{\hat H}:=E+{k \over 2}{\hat r}^{2}.\end{equation}

\vskip 10pt

Common eigenfunctions and eigenvalues of the maximal set of mutually
commuting hermitian differential operators $\hat s$, ${\hat J}_{(1)}$,
${\hat J}^{2}$, ${\hat H}$ may now be determined, in the usual
fashion, by the following set of differential equations:

\begin{equation}{\hat s}f(\pi_{A^{\prime}}, \ {\bar \pi}_{A})
= sf(\pi_{A^{\prime}}, \ {\bar \pi}_{A}),\end{equation}

\begin{equation}{\hat J}_{(1)}f(\pi_{A^{\prime}}, \ {\bar \pi}_{A})
= mf(\pi_{A^{\prime}}, \ {\bar \pi}_{A}),\end{equation}

\begin{equation}{\hat J}^{2}f(\pi_{A^{\prime}}, \ {\bar \pi}_{A})
= J(J+1)f(\pi_{A^{\prime}}, \ {\bar \pi}_{A}),\end{equation}

\begin{equation}{\hat H}f(\pi_{A^{\prime}}, \ {\bar \pi}_{A})
=\epsilon f(\pi_{A^{\prime}}, \ {\bar \pi}_{A}).\end{equation}

\vskip 10pt

We will be looking for a set of solutions of the above equations in the form:

\begin{equation}
f(\pi_{A^{\prime}}, \ {\bar \pi}_{A})
=f_{\sigma}(\pi_{A^{\prime}}, \ {\bar \pi}_{A})
Y_{ln}(\theta, \ \varphi)
f_{\epsilon}(E)
\end{equation}

\vskip 10pt

\noindent
where $f_{\sigma}(\pi_{A^{\prime}}, \ {\bar \pi}_{A})$ is any of the
following four holomorphic (either in ${\bar \pi}_{A}$ or in
$\pi_{A^{\prime}}$) simple expressions:

$$({\bar \alpha}^{A}{\bar \pi}_{A})^{-\sigma} \qquad
({\bar \beta}^{A}{\bar \pi}_{A})^{-\sigma} \qquad
(\alpha^{A^\prime}\pi_{A^\prime})^{-\sigma} \qquad
(\beta^{A^\prime}\pi_{A^\prime})^{-\sigma}.$$

\vskip 10pt

The analysis in the sequel does not depend on which of these four
options we assume so the second one is adopted for definitness.  We
just note that complex conjugation "creates" a massless particle with
reversed eigenvalue of the helicity operator while the two
remaining options correspond to the different choices of
the "direction" of the helicity relative to the "direction" of
the orbital magnetic quantum number $n$ (see 3.33).

\vskip 10pt

$\sigma$ is a positive integer (say) and $Y_{ln}(\theta, \
\varphi)$ represent the usual spherical harmonics where $\varphi$ and
$\theta$ are functions of the spinor variables as defined in
(3.11)-(3.12).  Spherical harmonics are thus homogenous functions of
degree $0$ in spinor variables.  Besides for any $f_{\epsilon}(E)$
where $E$ is given by (3.9) and $\hat s$ by (3.5) one has	:

\begin{equation}
{\hat s}f_{\epsilon}(E)=0.
\end{equation}

\vskip 10pt

Therefore we obtain:

$${\hat s}
f(\pi_{A^{\prime}}, \ {\bar \pi}_{A})
= {\hat s}[({\bar \beta}^{A}{\bar \pi}_{A})^{\sigma}
Y_{ln}(\theta, \ \varphi)f_{\epsilon}(E)]
=
Y_{ln}(\theta, \ \varphi)f_{\epsilon}(E)
{\hat s}({\bar \beta}^{A}{\bar \pi}_{A})^{\sigma}
=$$
\begin{equation}={\sigma \over 2}
({\bar \beta}^{A}{\bar \pi}_{A})^{\sigma}
Y_{ln}(\theta, \ \varphi)f_{\epsilon}(E)
=s f(\pi_{A^{\prime}}, \ {\bar \pi}_{A})\end{equation}

\vskip 10pt

\noindent
which shows that eigenvalues of the helicity operator are given by
$s={\sigma \over 2}$.

\vskip 10pt

Simple calculations also show that:

$${\hat J}_{(1)}
f(\pi_{A^{\prime}}, \ {\bar \pi}_{A}) =
Y_{ln}(\theta, \ \varphi)f_{\epsilon}(E)
{\hat J}_{(1)}({\bar \beta}^{A}{\bar \pi}_{A})^{-\sigma}
+
({\bar \beta}^{A}{\bar \pi}_{A})^{-\sigma}
f_{\epsilon}(E)
{\hat J}_{(1)}Y_{ln}(\theta, \ \varphi)$$

\begin{equation}=({\sigma \over 2} + n)
({\bar \beta}^{A}{\bar \pi}_{A})^{-\sigma}
Y_{ln}(\theta, \ \varphi)f_{\epsilon}(E)
= m f(\pi_{A^{\prime}}, \ {\bar \pi}_{A}).\end{equation}

\vskip 10pt

The eigenvalues of ${\hat J}_{(1)}$ are thus given by half-integer or
integer numbers:

\begin{equation}m = {\sigma \over 2} + n.\end{equation}

\vskip 10pt

{}From standard considerations (see e.g. Ref. 14) it now follows
that, for a given eigenvalue of the operator ${\hat J}^{2}$
the quantum numbers $m$ are bounded by certain upper and lower limits.
If we denote the upper limit by a positive number $J$, then
one obtains\cite{ll}
that the eigenvalue of ${\hat J}^{2}$ is $J(J+1)$. Note that
the positive number $J=l+{\sigma \over 2}$, where $l$ is the upper limit
(for a given eigenvalues of ${\hat J}^{2}$ and of $\hat s$)
of $n$ in (3.34), may assume half-integer or integer values.

\vskip 10pt

Finally we use the eigenvalue equation in (3.29) which, given quantum
numbers $s$, $m$ and $J$, determines the eigenfunctions $f_{\epsilon}$
and the corresponding eigenvalues of the total energy operator $\hat
H$ in (3.25). When written out explicitly the equation in (3.29)
reduces itself to an ordinary differential equation of second order:

\begin{equation}-{d^{2}f_{\epsilon} \over dE^{2}}
-{5 \over E}{df_{\epsilon} \over dE}
+({a \over E^{2}} + b E)f_{\epsilon} = \lambda f_{\epsilon},\end{equation}

\vskip 10pt

\noindent
where

\begin{equation}a=J(J+1) - ({\sigma^{2} \over 2} + 2\sigma +12),
\qquad b={2 \over k}, \qquad
\lambda={2\epsilon \over k}.\end{equation}

\vskip 10pt

The energy levels $\lambda$ and the corresponding eigenfunctions
$f_{\epsilon}$ may
be calculated  explicitly on a computer using
standard methods e.g.  the Ritz method\cite{asd}.  The already
mentioned computer program Maple V, Release 3 may be of great value
here.  The results of such explicit calculations we hope to be able to
present at a later occasion.

\section{\hspace{-4mm}\hspace{2mm}CONCLUSIONS AND REMARKS.}

If "elementary" particles such as e.g.  electron, proton, neutron etc.
may be regarded as bound states of a finite number of massless
spinning parts then twistor theory combined with the idea of
relativistic action at a distance should provide a very powerful tool
for construction of such models.

\vskip 10pt

There arises a possibility to use a twistor phase space formulation.
Such a reducible phase space may then be thought of as a direct
product of a finite number of copies of an elementary twistor phase
space \bf T.  \rm Dynamics will then be generated by an appropriately
chosen Poincar{\'e} scalar hamiltonian function.

\vskip 10pt

In this paper we therefore emphasized particle aspects of Penrose's
twistor formalism as opposed to the standard treatments where field
aspects are at the front.

\vskip 10pt

The suggested non-standard quantization in the previous section of
this note corresponds to the real polarization of the twistor phase
space as described by Wood- house\cite{woo}.  Choosing this non-standard
quantization we however loose some of the results of conventional
twistor theory such as the twistor description of massless free fields
in terms of holomorphic sheaf cohomology, the scalar product on such
fields, geometrization of the concept of positive frequency of the
field and the relationship between conformal curvature and the twistor
"position" (twistor variables) and "momentum" (complex conjugates of
the twistor variables) operators\cite{pr3}.

\vskip 10pt

What we gain is that the real dimension of the relativistic
configuration space of a massless spinning particle is one half of the
real dimension of the configuration space obtained by means of the
conventional holomorphic twistor quantization\cite{prmc}.  Further, the
configuration space obtained in our paper has a clear physical
interpretation.  Wave functions on such a configuration space define
quantum states in (the "square root" of) the linear momentum
representation.  However in our opinion the most important gain is the
fact that using our formulation we are able to treat \it interacting
\rm massless spinning particles (not fields) forming a closed composite
bound system.

\vskip 10pt

Using ideas presented in this paper and in\cite{ab1} it would be
interesting to investigate a fully relativistic closed system forming
a massive and spinning particle composed of three or four directly interacting
massless and spinning parts.

\vskip 10pt

Similar in spirit, attempts to remodel the physics of elementary
particles, have been made before by Hughston\cite{hl},
Popovich\cite{pa} and Perj{\`e}s\cite{pz}.  These authors make use of
conventional twistor quantization and as primary objects regard \it
free \rm fields which are then represented by elements of the
holomorphic sheaf cohomology group of an appropriate twistor space.

\vskip 10pt

Finally we note that commutation relation in (2.36) may have an
experimental implication. The uncertainty principle following from
(2.36) predicts that position of a massless and spinning particle can
never be measured exactly. No sharp value of its position vector
exists.

\vskip 10pt

The speculations presented in this note are inconclusive with respect
to their physical significance.  Some calculations are in progress in
order to find phenomenological support for the presented ideas.

\vskip 10pt

Nevertheless, our attempts seem to comply with the Twistor
Programme announced by Penrose\cite{penr}.


\begin{thebibliography}{99}

\bibitem{shm}  Schwartz H.M. (1968), "Introduction to Special relativity"
p. 123, McGraw-Hill.

\bibitem{prmc} Penrose R. and MacCallum M.A.H. (1972), Physics report C6(4),
241.

\bibitem{wfw} Kerner E.H. (Editor) (1972), "The theory of action-at-a-distance
in relativistic particle dynamics - reprint collection", Gordon and Breach,
Science Publishers, Inc.

\bibitem{ab1} Bette A., (1984), J.  Math.  Phys.  Vol.  25.  (the Poisson
bracket relation in (2.22) and the following conclusions although
never made use of are incorrect and should be replaced by the Poisson
bracket relation corresponding to (2.36) in the present paper.)

\bibitem{ab2}
Bette A., (1988), in "Microphysical Reality and Quantum formalism"
pp.  447 - 455., A.  Van der Merwe et al.  (eds.), Kluwer Academic
Publishers.

\bibitem{rw}  Rindler W. (1982), "Introduction to Special relativity" formula
(35.12) p.104, Oxford University Press.

\bibitem{pr1} Penrose R.  (1968), in "Batelle Rencontres" eds.  C.M.  de Witt,
J.A. Wheeler Princeton University W.A.  Benjamin inc.  New York,
Amsterdam pp.135-149.

\bibitem{prrw} Penrose R.  and Rindler W.  (1984), "Spinors and Space-Time",
Cambridge Monographs on Mathematical Physics vol.1 and 2, Cambridge
University Press.

\bibitem{pr2} Penrose R.  (1972), in "Magic without magic: J.  A.  Wheeler a
collection of essays in honor of his 60th birthday.", ed.  J.R.
Klauder, W.H.  Freeman and Co., San Francisco.

\bibitem{hl} Hughston L.P.  (1979), "Twistors and Particles", Lectures notes
in Physics No.97, Springer Verlag, Berlin, Heidelberg, New York.

\bibitem{ab} Bette A.  (1994), "Hamiltonian Dynamics of Massless Objects",
in "Proceedings of the XII Workshop On Geometric Methods in Physics",
Editors: S.Twareque Ali, I.M. Mladenov, A. Odzijewicz. Plenum Press.

\bibitem{dpam1} Dirac P.A.M. (1925), Proc. Roy. Soc. London ser. A, 109,
642-653.

\bibitem{dpam2} Dirac P.A.M.  (1958), The Principles of Quantum Mechanics,
Fourth
edition, Oxford, Clarendon Press.

\bibitem{ll} Landau L.D.  and Lifshitz E.M., (1975), "Quantum Mechanics - non -
relativistic theory" Fourth Revised English Edition, Pergamon Press,
Addison-Wesley Publishing Company Inc.

\bibitem{asd} Davydov A.S., (1963) "Kvantovaja Mechanika",
sect. 51, "Gosudarstvennoje Izdatelstvo
Fiziko-Matematitjeskoj Literatury", Moskva, (in Russian).




\bibitem{woo} Woodhouse N.  (1992), "Geometric Quantization" -
Second edition in
Oxford Mathematical Monographs series, Oxford Ueries, Oxford University Press,
Clarendon Press-Oxford, p.126 and p.193.

\bibitem{pr3} Penrose R. (1968), Int.J.Theor.Phys., Vol.1, No.1, 61-99.

\bibitem{pa} Popovich A.S., (1980), Doctoral thesis at Oxford's Mathematical
Institute.

\bibitem{pz} Perj{\`e}s Z., (1979), Phys. Rev. D, Vol.20, No. 8.

\bibitem{penr} Penrose R. (1977), Rep.Math.Phys., Vol.12, p.65.
\end{thebibliography}
\end{document}